\documentclass[12pt,letterpaper]{article}
\pdfoutput=1

\usepackage{scicite}
\usepackage{times}

\usepackage[pass]{geometry}
\usepackage{graphicx}
\usepackage{listings}
\usepackage{xcolor}
\usepackage{booktabs}
\usepackage{url}
\usepackage{rotating}
\usepackage{xr}

\let\long\relax

\ifx\review\relax
\fi

\newenvironment{sciabstract}{%
\begin{quote}
\end{quote}}

\graphicspath{{.}{../../../src/research/alexandria3k/docs/schema}}

\definecolor{background}{HTML}{EEEEEE}
\definecolor{dkgreen}{rgb}{0,0.6,0}
\definecolor{gray}{rgb}{0.5,0.5,0.5}
\definecolor{mauve}{rgb}{0.58,0,0.82}

\ExplSyntaxOn 
\cs_new:Npn \expandableinput #1
  { \use:c { @@input } { \file_full_name:n {#1} } }
\ExplSyntaxOff

\title{Open reproducible publication research}

\author{Diomidis Spinellis$^{1,2\ifx\review\relax\ast\fi}$\\
\\
\normalsize{$^{1}$Department of Management Science and Technology,}\\
\normalsize{Athens University of Economics and Business,}\\
\normalsize{Athens, Greece}\\
\normalsize{$^{2}$Department of Software Technology,}\\
\normalsize{Delft University of Technology,}\\
\normalsize{Delft, The Netherlands}\\
\\
\ifx\review\relax
\normalsize{$^\ast$To whom correspondence should be addressed; E-mail:  dds@aueb.gr.}
\fi
}

\date{}

\begin{document}

\newcommand{\surl}[1]{\small{\url{#1}}}
\newcommand{\name}{Alexandria3k}
\newcommand{\cdv}{CD\textsubscript{5}}

\maketitle

\subsection*{Abstract}
\begin{sciabstract} 
Considerable scientific work involves locating, analyzing, systematizing, and
synthesizing other publications, often with the help of
online scientific publication databases and search engines,
However, use of online sources suffers
from a lack of repeatability and transparency,
as well as from technical restrictions.
\name\ is a Python software package and an associated command-line tool
that can populate embedded relational databases
with slices from the complete set of several open publication metadata sets
for reproducible processing through sophisticated and performant queries.
We demonstrate the software's utility
by visualizing the evolution of publications in diverse scientific
fields and relationships among them,
by outlining scientometric facts associated with COVID-19 research, and
by replicating commonly-used bibliometric measures of productivity, impact,
and disruption.
\end{sciabstract}

\ifx\review\relax
\subsection*{Teaser}
\begin{quote}
Alexandria3k and open data allow the performance of sophisticated and reproducible publication studies on a personal computer.
\end{quote}
\fi

\section*{Introduction} 
\begin{figure}[!t]
\includegraphics[width=\columnwidth]{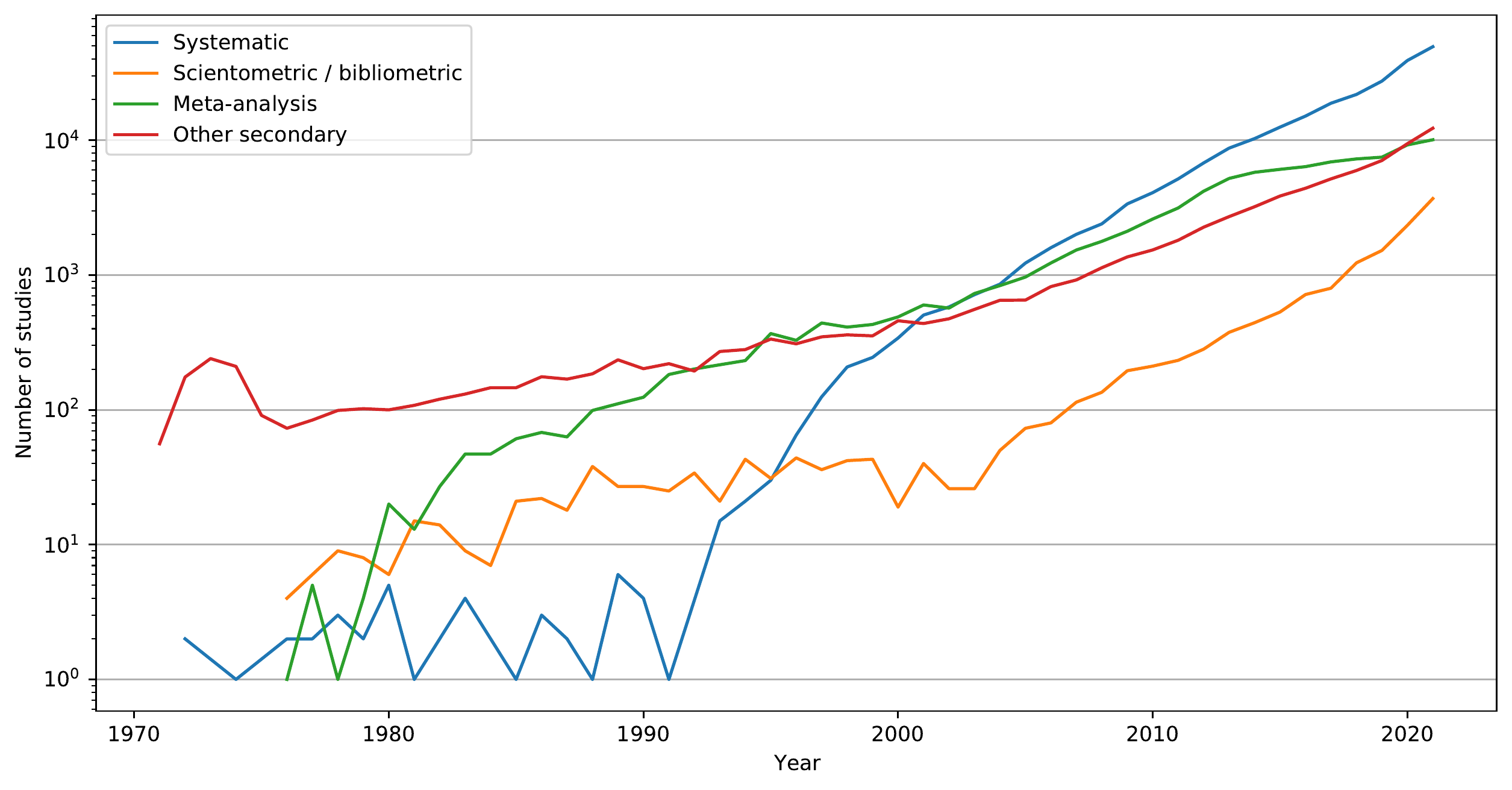}
\caption{\label{fig:synthesis}Number of research synthesis studies published
each year in the period 1971--2021.}
\end{figure}

Research synthesis is becoming an increasingly important\cite{GKNS18}
and popular scientific method.
By our own calculations
about 437 thousand scientific studies published from 1846 ---
the year of the first one we found\cite{BBF46} --- onward
are based on the analysis of previously published primary studies.
(The Methods section provides details on how
all numbers appearing in this report were obtained in a repeatable manner.)
\ifx\long\relax
Synthesizing studies are typically identified by their titles through terms such as:
``systematic review'', ``systematic literature review'', or
``systematic mapping study''
  (secondary studies using methods that help make their findings unbiased
  and repeatable --- 251\,850 titles);
``secondary study'', ``literature survey'', or ``literature review''
  (a not necessarily systematic study reviewing primary studies
  --- 77\,037 titles);
``tertiary study'' or ``umbrella review''\cite{AFGH15}
(a study reviewing secondary studies --- 4\,039 titles);
``meta-analysis'' (a systematic secondary study employing statistical methods --- 92\,363 titles);\cite{KBB15}
as well as (systematic by definition)
``scientometric'' (employing quantitative methods to study scientific research --- 2\,769 titles)
and ``bibliometric'' (using statistical methods to study written communications --- 12\,361 titles)
studies\cite{KBB15}.
The number of synthesis studies
\else
The number of synthesis studies~\cite{AFGH15,KBB15}
\fi
published each year has risen considerably over the past two decades,
particularly for systematic literature reviews
(see Figure \ref{fig:synthesis}).
A major data source for research synthesis studies
are online specialized and general purpose
bibliographic and article databases\cite{Fin10},
such as
ERIC,
Google Scholar,
Inspec,
Scopus, and
Web of Science.

Performing systematic studies on published literature through the available
online systems can be problematic\cite{GH20}.
First, their
constantly updated contents,
bubble effects\cite{CK18},
location-- and license-dependent results\cite{PLVG21},
and periodic changes to their internal workings
compromise reproducibility\cite{KR16,HCCK17}.
Even when the search strategy is well-documented to aid reproducibility
by following recommended reporting guidelines such as PRISMA\cite{PMBB21},
which is often not the case\cite{YPKA09,MTK11},
it is difficult to repeat a query to an online service, and
obtain the same results as those that have been published\cite{GH20}.
In addition, the reproducibility of such studies is hampered
in the short term by the fees required for accessing some online services, and
in the long term by the commercial survival
of the corresponding companies\cite{Fri04}.
Service access costs on their own can restrict institutions
with limited funding from conducting systematic literature studies.
Another associated problem is the lack of transparency\cite{Noo14}.
Most online services work with proprietary data collections and algorithms,
making it difficult to understand and explain the obtained results.
As an example, Clarivate's journal impact factor calculation depends
on an opaque collection of journals\cite{Fer16}
and list of ``citable items''\cite{PME06} tagged so by the vendor.
Finally, there are technical limitations.
Some services lack a way to access them programmatically
(an application programming interface --- API)\cite{GH20},
forcing researchers to resort to tricky and unreliable contortions,
such as screen scrapping.
Both APIs and offered query languages are not standardized\cite{SMCG09},
and often restrict the allowed operations\cite{GH20}.
For instance, the network-based APIs suffer
from corresponding latency\cite{BW16}, and often
from rate and ceiling limits to the number of allowed invocations\cite{BSPB21}.
These restrictions hinder studies requiring
a large number or sophisticated queries.

The outlined problems can be addressed thanks to
sustained
exponential advances in computing power\cite{Sch97},
drops in associated costs, and
Open Science initiatives\cite{NABB15}.
The \name\ system presented here
is an open-source software library and command-line tool
that builds on these advances to allow the conduct of
sophisticated systematic research of published literature,
(e.g. literature reviews, meta-analyses, bibliometric and scientometric studies)
in a
transparent,
repeatable,
reproducible, and
efficient
manner.
\name\ allows researchers to process on a personal computer
publications' metadata (including citations)
from most major international academic publishers
as well as corresponding author, funder, organization, and journal details.
Specifically, \name\ works on data snapshots offered periodically by
initiatives, such as
Crossref (publication metadata, journal names, funder names)\cite{Lam15},
ORCID (author details)\cite{HFPP12},
ROR (research organization registry)\cite{Lam20},
and others.
Using \name\ researchers can query and process these data through
SQL queries launched by means of
command-line tool invocations or Python scripts.
Researchers can ensure the transparency, reproducibility,
and exact repeatability of their methods
by documenting or publishing (when permitted)
the version of the data used and the employed
commands\cite{NABB15}.
The primary data sets can be stored and processed locally on a modern
laptop, because they amount to a few hundred gigabytes in their compressed
format (157 GB for Crossref, 25 GB for ORCID data;
the downloading of the Crossref data is facilitated by its availability
through the BitTorrent protocol)\cite{AS04}.
The data are decompressed in small chunks ensuring that both main and
secondary memory requirements are kept within the limits of what
a typical personal computer can accommodate.
(Keeping the data decompressed or populating a relational or graph database
with all of it would require more than 1.5 TB of storage space.)
In addition, \name\ offers facilities for
running relational database queries on data partitions,
sampling records, and
populating a relational database with
a subset of records or fields.
All these facilities help tasks to execute in reasonable time.
On a populated and suitably indexed database with millions of records
many queries finish in minutes.
Queries or database population tasks
involving a full scan of the entire Crossref publication data set
complete in about five hours.
(Full performance details are provided in the Methods section.)

\section*{Contents, structure, and use} 
In total, \name\ offers relational query access to 2.6 billion records.
These are organized in a relational schema illustrated in 
Figures~\ref{fig:schema-crossref}--\ref{fig:schema-other}.

\begin{table}
\caption{\label{tab:crossref}Number of Crossref Records}
\begin{center}
\begin{tabular}{lr}
\toprule
Entity & Records \\
\midrule
Total records	& 2\,531\,227\,295 \\[.7ex]

Works (publications)	& 134\,048\,223 \\
Works with a text mining link	& 96\,294\,821 \\
Works with subject	& 81\,210\,089 \\
Works with references	& 52\,907\,361 \\
Works with affiliation	& 16\,833\,863 \\
Works with an abstract	& 15\,367\,820 \\
Works with funders	& 7\,519\,462 \\[.7ex]

Author records (linked to works)	& 359\,556\,891 \\
Author records with ORCID	& 16\,745\,506 \\
Distinct authors with ORCID	& 4\,525\,906 \\[.7ex]

Author affiliation records	& 76\,768\,648 \\
Distinct affiliation names	& 19\,453\,360 \\[.7ex]

Work subject records	& 182\,858\,177 \\
Distinct subject names	& 340 \\[.7ex]

Work funders	& 15\,491\,915 \\
Funder records with DOI	& 10\,811\,496 \\
Distinct funder DOIs	& 29\,610 \\
Funder awards	& 14\,090\,597 \\[.7ex]

References 	& 1\,748\,421\,617 \\
References with DOI	& 1\,255\,033\,889 \\
Distinct reference DOIs	& 59\,127\,679 \\
\bottomrule
\end{tabular}
\end{center}
\end{table}

\begin{figure}
\includegraphics[width=\columnwidth]{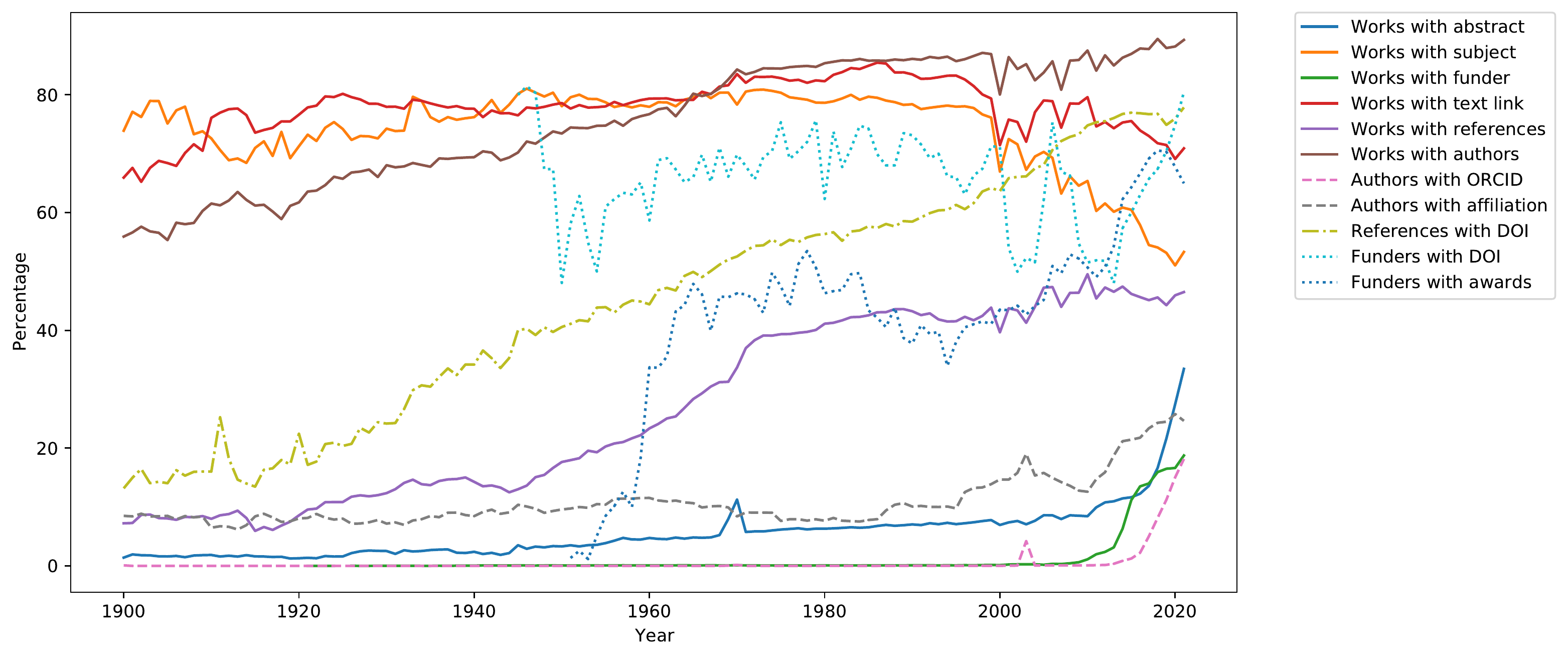}
\caption{\label{fig:yearly-availability}Availability of
Crossref data elements over the period 1900--2021.
Text link availability refers to full-text mining links.
Availability of author ORCID and affiliations
is evaluated over all of each year's individual
author records
appearing alongside publications.
Similarly, availability of funder DOI and award data
is evaluated over all of each year's individual
funder records
appearing alongside publications.
This means e.g. that if an author appears in one publication with an ORCID
and in another without,
the reported ORCID availability for the corresponding year will be 50\%.
Availability of funder, text link, references, authors, affiliation, and awards
refers to the existence of at least one such record associated with the parent
entity.
DOI availability in references and funders is evaluated for each individual
record.
}
\end{figure}

Most records are publication metadata obtained from the
Crossref Public Data File.
These contain
publication details
(DOI, title, abstract, date, venue, type, pages, volume, ...),
a publication's references to other publications
(DOI, title, author, page, ISSN, ISBN, year, ...),
and other data associated with each publication's
authors and their affiliations,
funders and funder awards,
updates (e.g. retractions),
subjects,
licenses, and
hyperlinks for text mining of the publication's full text\cite{Lam15}.
Details about the data available through Crossref are listed in
Table~\ref{tab:crossref}.
Note that coverage is
\ifx\long\relax
incomplete;
for example,
39\% of the publications have a reference list associated with them,
70\% of funders are uniquely identified with a DOI,
while only 11\% of the publications have an abstract.
For most types of records coverage is generally increasing over time
\else
incomplete
\fi
(see Figure~\ref{fig:yearly-availability}).

\begin{table}
\caption{\label{tab:orcid}Number of ORCID Records}
\begin{center}
\begin{tabular}{lrr}
\toprule
Table	& 			& Persons with  \\
	& Records		& Such Records \\
\midrule
Personal data	& 14\,811\,567	& 14\,811\,567 \\
URLs	& 1\,325\,399	& 892\,528 \\
Countries	& 2\,141\,021	& 2\,021\,218 \\
Keywords	& 2\,230\,569	& 838\,796 \\
External identifiers	& 2\,287\,063	& 1\,653\,995 \\
Distinctions	& 345\,967	& 159\,113 \\
Educations	& 5\,610\,051	& 3\,087\,381 \\
Employments	& 5\,932\,529	& 3\,472\,114 \\
Invited positions	& 263\,315	& 147\,497 \\
Memberships	& 660\,304	& 357\,799 \\
Qualifications	& 940\,400	& 585\,298 \\
Services	& 221\,860	& 108\,288 \\
Fundings	& 1\,170\,432	& 328\,958 \\
Peer reviews	& 6\,270\,131	& 751\,404 \\
Research resources	& 3\,012	& 1\,779 \\
Person's works	& 34\,237\,877	& 3\,224\,068 \\
\midrule
Total	& 78\,451\,497	&  \\
\bottomrule
\end{tabular}
\end{center}
\end{table}

\name\ can unambiguously link Crossref records to imported author metadata
through ORCID (Open Researcher and Contributor ID):
a non-proprietary system developed to identify authors
in scholarly communication\cite{HFPP12}.
ORCID tables that \name\ supports include those detailing
an author's
URLs,
countries,
keywords,
external identifiers,
distinctions,
education,
employment,
invited positions,
memberships,
qualifications,
services,
fundings,
peer reviews,
used research resources,
and published works.
Most of these tables contain
details of the associated organization
(name, department, city, region, country),
the role title,
and the starting and end date.
The currently available ORCID data set contains about 78 million records
associated with 14 million authors.
The completeness of the ORCID records is low and uneven
(see Table~\ref{tab:orcid}),
which means that research based on it must be carefully designed.

\name\ can also import the Research Organization Registry data\cite{Lam20}
containing details of 104\,402 organizations, as well as related
acronyms (43\,862 records) and
aliases (25\,119 records).
Through the provided ROR identifier it can
link unambiguously elements from
a person's employment and education ORCID records to the corresponding
organization.
Currently ORCID contains such identifiers for
130\,033 employment records and
133\,066 education records.
Given that
only 4.6\% of work author records have an ORCID and
only 23\% of ORCID records contain employment information,
\name\ also provides a performant facility to match the textual affiliation
information listed in works and link it to ROR identifiers.

Finally, \name\ can import and link three reference tables:
the names of journals associated with ISSNs (currently 109\,971 records),
the funder names associated with funder DOIs (32\,198 records), and
the metadata of open access journals (18\,717)\cite{Mor17}.
\name\ further disaggregates journal ISSNs according to their type
(electronic, print, or alternative --- 158\,580 records).

The data used by \name\ is openly distributed by diverse third parties
(see the data availability statement)
in textual tree or flat format files:
JSON for Crossref and ROR, XML for ORCID, and CSV for the rest.
\name\ structures the data it offers in a relational schema
of 45 tables
linked through 47 relations.
Stored in a relational database and combined with suitable indexes,
this allows performing sophisticated analyses via SQL
(structured query language) queries in an efficient manner.
Records between diverse data sets are linked through standardized
globally unique identifiers:
DOIs for published works and funders,
ISSNs for journals,
ORCIDs for authors, and
RORs for research organizations.

\name\ is distributed as open source software in the form of a Python package,
which can be easily installed through the PyPI repository.
It can be used either as a command-line tool,
where its operation (e.g. query to run) is specified through
command-line arguments, or as a Python library,
which can be used interactively (e.g. by developing a
Jupyter Notebook\cite{KRBF16}) or through scripts.
The performance of both interfaces is similar,
because in both cases the heavy lifting is done by the same Python library code.

The latest version of the Crossref data is distributed as 26 thousand
compressed container files, each containing details about 5\,000 works.
A complete import of the Crossref data would amount to a 520 GB database.
Given the large amount of Crossref data, the population of a
database with it can be controlled in three ways.
First, only a horizontal subset of records can be imported,
by specifying an SQL expression that will select a publication
record only when it evaluates to {\tt TRUE}
(e.g. {\tt published\_year BETWEEN 2017 AND 2021}).
To facilitate the selection of records selected through other
means, the expression can also refer to tables of external databases.
Second, only a subset of the Crossref 26\,810 data containers
can be processed for import,
by specifying a Python function that will import a container when it
evaluates to {\tt True}.
This is mostly useful for random sampling, e.g. using
{\tt random.random() < 0.01} to sample approximately 1\% of the containers.
(A fixed seed value is used internally
for initializing the pseudo-random number generator
to allow deterministic and therefore repeatable sampling.)
Third, the populated tables or columns of the Crossref data set can be
vertically restricted by using the
{\tt table-name.column-name} or
{\tt table-name.*} SQL notation.
The population of a database with ORCID data can be also
horizontally restricted
to records associated with existing Crossref authors or published works
(probably selected in a previous population step)
and vertically restricted to include only specific tables or
columns, as in the case of Crossref.
Given their small volume,
no population controls are supported for the other data sets.

\section*{Application examples} 
The following paragraphs outline some simple
proof-of-concept applications of \name,
which demonstrate its use and motivate its adoption.
All are exactly replicable through
SQL queries and relational online analytical processing (ROLAP)
Makefiles\cite{MAKE,GS17,GS17b}
provided in the accompanying materials.

\begin{figure}
\includegraphics[width=\columnwidth]{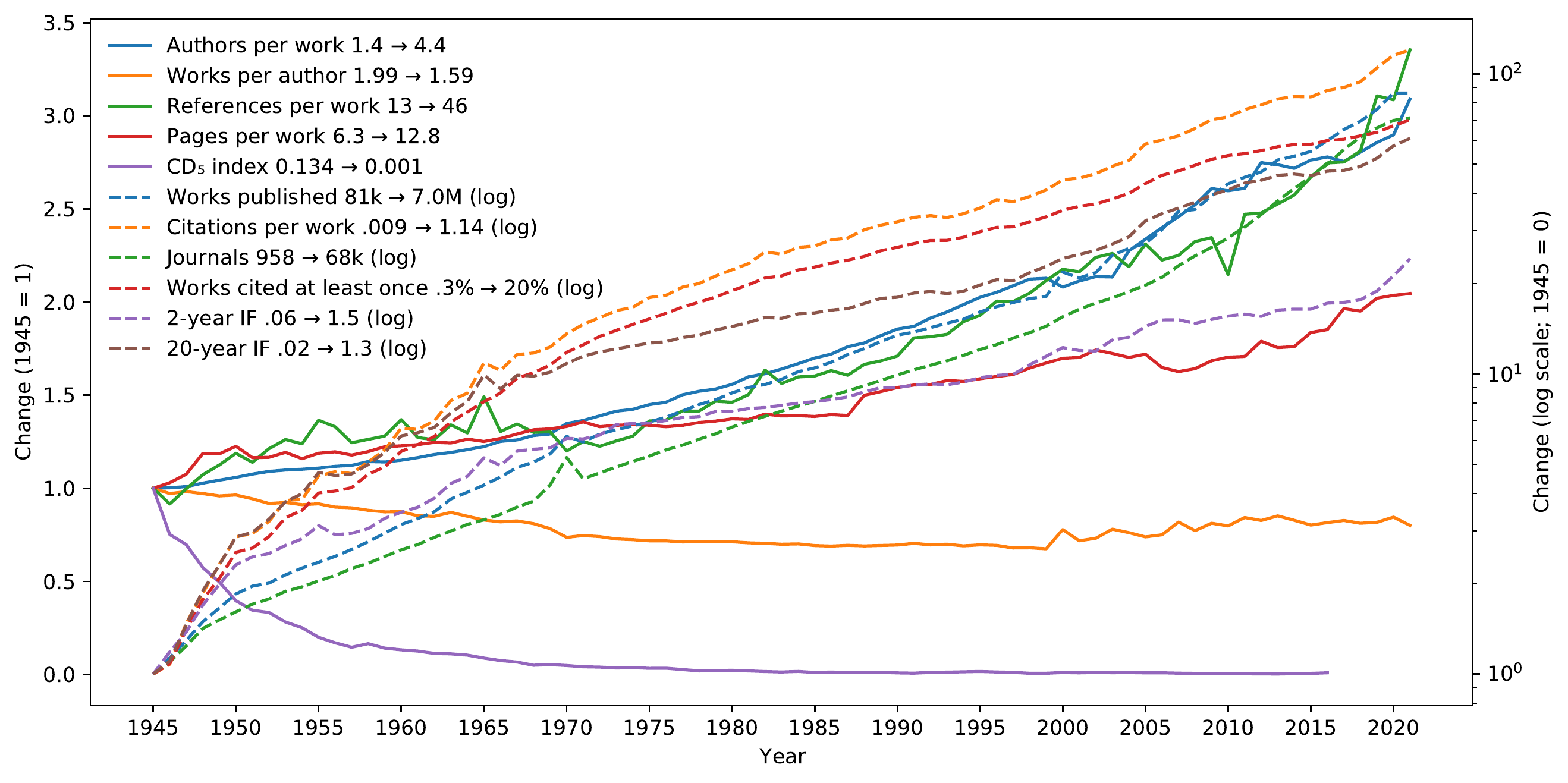}
\caption{\label{fig:yearly-evolution}Evolution of scientific publishing metrics in the post-WW2 period}
\end{figure}
Figure~\ref{fig:yearly-evolution}
showcases the use of \name\ to chart a view of
scientific publishing evolution in the post-WW2 period.
Despite the exponential increase in the number of published works
(accommodated by a corresponding swell in available journals),
publications are becoming ever more connected by citing each other.
This can be seen in
the rises of the references each work contains,
the citations works receive,
the phenomenal proportion of all works ever published that are
cited at least once every year (20\%),
and corresponding rises to the 2-year and even 20-year global impact factor.
Authors appear to be collaborating more and on longer papers
with only a slight decrease in the mean number that they publish each year.
The fall in the consolidation/destabilization (CD) index
is in line with recently published research reporting
that papers are becoming less disruptive over time\cite{PLF23}.
There is significant correlation
(Spearman rank-order correlation coefficient 0.95;
p-value $6\times10^{-34}$)
between the \cdv\ index
yearly averages obtained using \name\ with those available
in the previously published dataset,
which was obtained from data that are not openly available.
The sharp inflections in the Figure probably stem from artifacts
of the underlying data set,
and indicate that obtaining scientifically robust results would require
deeper analysis of the data.

\begin{figure}
\includegraphics[width=\columnwidth]{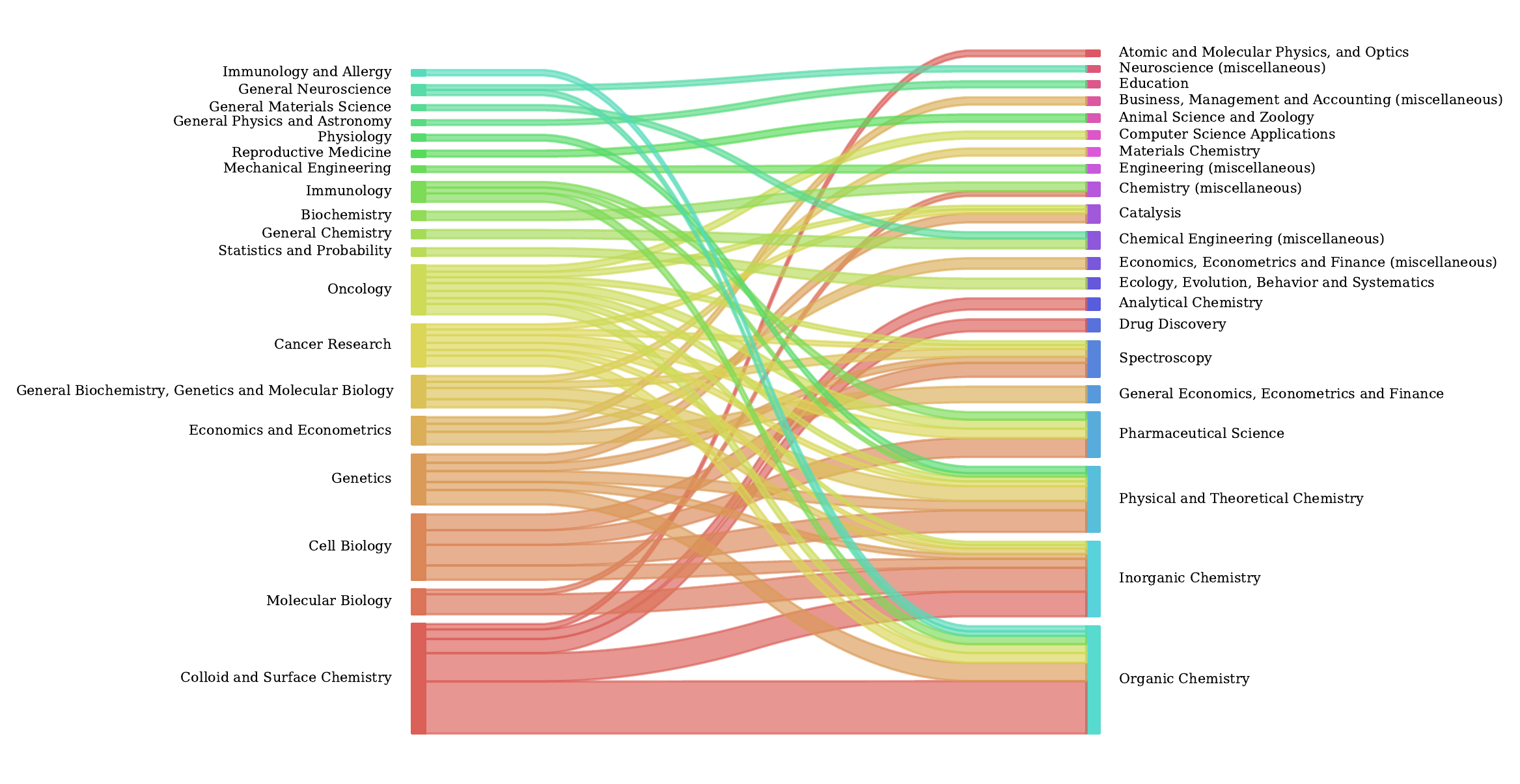}
\caption{\label{fig:subject-hierarchy}Strong relationships between citing and cited subjects}
\end{figure}
Linking publications with their specific scientific field and
their citations allows us to examine the
structure of the corresponding graph.
We can find fields that strongly depend on others by defining the
citation imbalance between two fields as
the ratio of one field's outgoing citations over the total
number of citations between them.
Figure~\ref{fig:subject-hierarchy} shows the fifty strongest field
relationships in terms of imbalance.
We note citing fields from the health and life sciences,
numerous cited fields associated with chemistry,
and the large number of fields
from which Oncology and Cancer Research draw upon.

\begin{figure}
\includegraphics[width=\columnwidth]{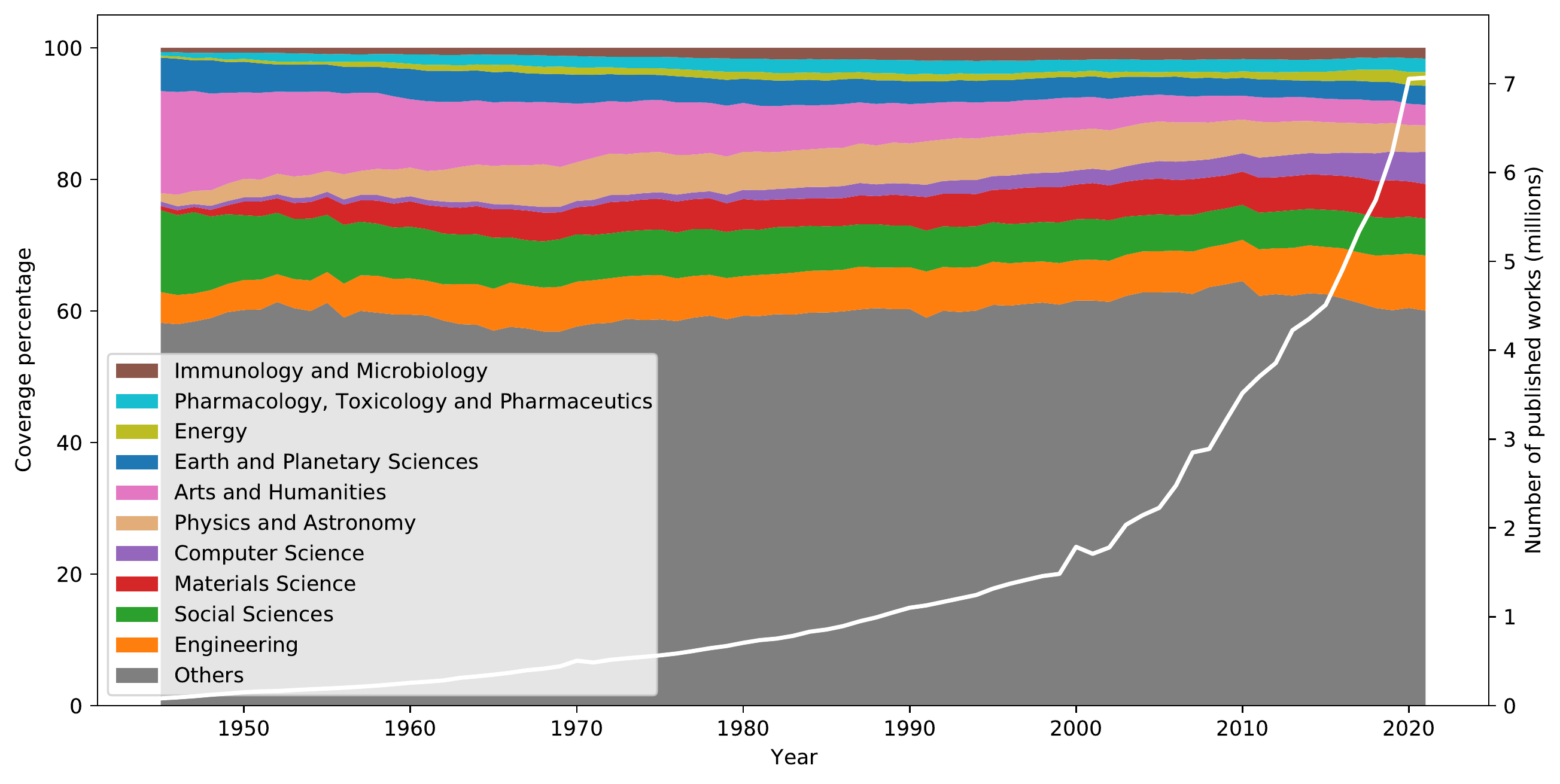}
\caption{\label{fig:fields-by-year}Evolution of subject coverage and publications 1945--2021}
\end{figure}
Associating publications with the general scientific field of
the journal they were published
(according to the Scopus All Science Journal Classification Codes --- ASJCs)
provides us a view of the evolution of the ASJC 27 fields over the years.
Figure~\ref{fig:fields-by-year} shows the ten fields amounting to more than
2\% of publications in 2021 that had the largest change in their publication
number in the period 1945--2021.
Clearly visible is the expected rise in Computer Science, Materials Science,
and Physics and Astronomy,
as well as the fall of Arts and Humanities and Social Sciences
publications.
Owing to the exponential rise of published works
(tallied on the Figure's right-hand axis)
the falls are only in relative terms:
2021 saw the publication of 291\,366 Arts and Humanities articles against
19\,873 in 1945.

\begin{figure}
\includegraphics[width=\columnwidth]{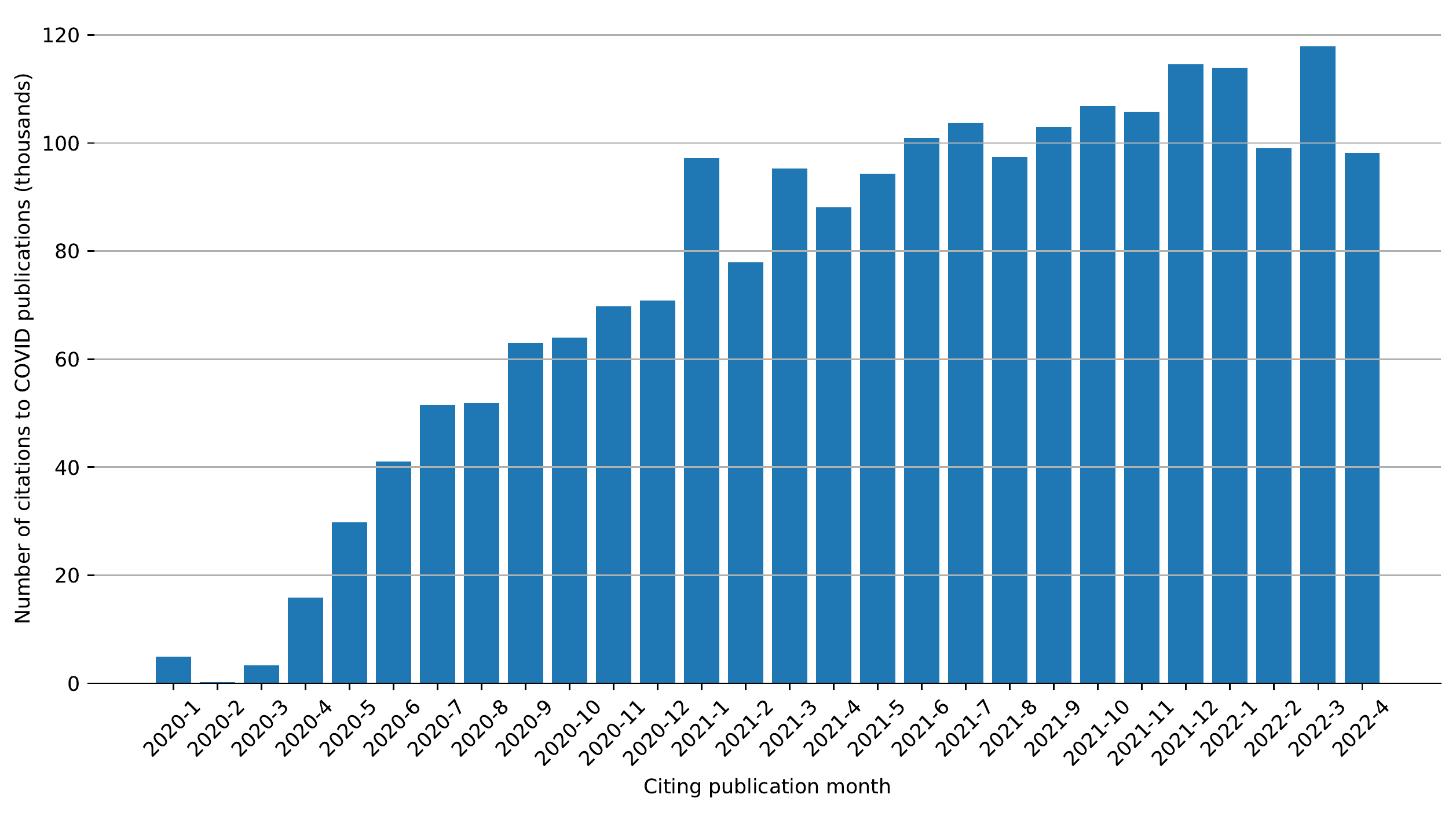}
\caption{\label{fig:covid-citations-time}Citations from COVID research to COVID research over time}
\end{figure}
We also looked how long it took for COVID articles to start citing
each other,
building, as it where, ``on shoulders of giants''.
As can be seen in Figure~\ref{fig:covid-citations-time}
a citation network ramped up relatively quickly,
surpassing ten thousand citations to COVID research by April 2020,
and reaching 118 thousand citations on March 2022.
The large number of works published in January 2020
appears to be due to journals publishing later in the year
volumes with that date.
This was, for example, the case with
an overview article on COVID-19 and cerebrovascular diseases\cite{TPZL20} and
an eight month retrospective\cite{AK20}.
(The same phenomenon may also explain the January 2021 rise.)
Notably, this backdating practice can distort
the establishment of priority over scientific advances
based on a journal's publication date.

To showcase how \name\ could be used to study
research related to COVID-19\cite{CAL20}
we examined publications containing COVID in their title or abstract.
We counted 491\,945 publications from about 1.5 million authors.
These covered 331 different topics demonstrating
the many disciplines associated with the research.
Some noteworthy topics and work numbers among
those with more than one thousand publications include
General Medicine (rank 1 --- 70\,609 works),
Psychiatry and Mental health (rank 4 --- 10\,404 works),
Education (rank 5 --- 9\,590 works),
Computer Science Applications (rank 18 --- 6\,013 works),
General Engineering (rank 20 --- 5\,942 works),
Strategy and Management (rank 42 --- 3\,208 works),
Law (rank 57 --- 2\,557 works),
History (rank 62 --- 2\,329 works),
Cultural Studies (rank 76 --- 1\,893 works),
Pollution (rank 97 --- 1\,549 works), and
Anthropology (rank 130 --- 1\,032 works).
Looking at listed funders, we saw that the top three in terms
of associated publications were
the National Natural Science Foundation of China (3\,506 works),
followed by the
(US) National Institutes of Health (2\,316),
the (US) National Science Foundation (1\,022),
the Wellcome Trust (914),
and the (UK) National Institute for Health Research (661).
We also examined the affiliations of COVID study authors,
propagating them to the highest parent organization
(e.g. a university hospital to its university).
Through this measure the top five entities were
the Government of the United States of America (1\,465 works),
the University of California System (925),
University of Toronto (910),
University of London (824), and
University of Oxford (660).

We replicated diverse impact and productivity metrics typically
calculated in a proprietary fashion by commercial bibliometric
data providers.
We calculated the 2021 Journal Impact Factor (JIF), and
compared the 600 journals with the highest JIF available through
Clarivate with 581 we matched through their ISSNs,
obtaining a Spearman rank-order correlation coefficient of 0.72
with a p-value $3\times10^{-93}$.
We also calculated the journal h5-index, and compared the values of
the 100 ``Top publications'' venues listed in Google Scholar\cite{Goo23all}
against those of \name\ by hand-matching the venue titles.
The comparison of the 91 matched elements gave
a Spearman rank-order correlation coefficient of 0.75
with a p-value $6\times10^{-18}$.
Examining the same metric in a field where considerable work is published
in conferences,
we compared the h5-index of the 13 common venues between the 20
Google Scholar reports in the ``Software Systems'' category\cite{Goo23ss}
against a curated list of 32 software engineering venues\cite{MAM19},
and obtained a Spearman rank-order correlation coefficient of 0.83
with a p-value $0.0004$.

Through the same data we also obtained the most cited articles
published in the corresponding period
(predictably, the top-one, with 21\,426 citations, was on COVID-19)\cite{HWXR20}
and overall
(the top one was a 1996 article on
the generalized gradient approximation method\cite{Per96},
which received 39\,715 citations from a surprising diversity of fields).

We also obtained the h5-index on authors identified through ORCID.
A noteworthy observation was the number of authors with a high metric:
the top-ranked author has an h5-index of 76,
twelve authors have an h5-index larger than 60, and
100 larger than 38.
These achievements appear to be even more difficult to explain
than earlier observations of hyperprolific authors\cite{IKB18}.
We explored the phenomenal productivity and impact exhibited
by the authors at the top of the distribution by examining
the clustering coefficient of the graph induced by incoming and outgoing
citations of distance 2 for a given work.
The clustering coefficients of a random sample of 50 works from authors
with an h5-index larger than 50 (median coefficient 0.05)
appear to be significantly different
from a random sample of other works of the same size
with the same number of citations for each one (median 0.03).
Specifically, comparing the two populations we obtained
a Mann-Whitney U statistic measure of 781 for the coefficients of the
top-ranked authors' works with a p-value of 0.0006.

\section*{Discussion} 
\name\ builds on
a rich and evolving ecosystem of publicly available data and
sophisticated open source software libraries coupled with exponential
advances in computing power,
enabling scientists to perform
reproducible bibliometric, scientometric, and research synthesis studies
in a transparent and repeatable manner
on a personal computer.
In our use of \name\ we found that
over time its features and interface, as well as
the methods we used to conduct the proof-of-concept studies,
crystallized into a form we employed for distributing the code
of all the proof-of-concept studies.
Other researchers can readily build upon these examples
to conduct their own studies.

\name\ is not a panacea for the issues we identified in the
opening paragraphs.
Crossref's coverage of citation links is lacking compared to
commercial alternatives\cite{VvW21}.
The linkage of publications to their authors through ORCID is thin,
and ORCID author metadata is also sparse.
The string-based matching used to link textual author affiliations
to RORs can result in mismatches.
The subjects associated with works are derived from the categorization of
whole journals by Scopus,
and they are therefore incomplete and potentially inaccurate.
The absolute values of derived bibliometric measures do not match
the proprietary ones, which employ different methods
for data collection (web crawling for Google Scholar) and
for processing (hand-curation for Clarivate).
In general, many of the supported data elements can be used for making
statistical observations,
but the validity of any findings needs to be verified,
e.g. through sensitivity analysis based on the manual examination of a sample.
Finally, one should also take into account the epistemological
shortcoming of structured approaches that use publication databases\cite{BC10}.

Nevertheless, \name\
opens many possibilities to conduct research that goes beyond
what can be easily done through the query and API-based approaches offered
by existing proprietary\cite{Fin10} and open\cite{PPO22} databases.
Examples include the study:
of collaboration patterns between organizations and disciplines,
of citation cliques and organizational inbreeding,
of funder, publisher, and organizational performance,
of open access availability and its effects,
of pre-print servers as alternatives to peer-reviewed publications,
of interdisciplinary connections, and
of structural publication differences between organizations or
scientific fields.

Apart from making \name\ available as open-source software we also
structured it and intend to run it as an open-source project,
accepting and integrating contributions
of additional open-data sources (e.g. MEDLINE/PubMed, patent metadata),
algorithms (e.g. publication topic classification;
author name and affiliations disambiguation;
citation matching), and example studies.
We hope that this will allow \name\ to grow organically
serving ever more needs of research synthesis and analysis studies.

\section*{Methods} 
The paragraphs below describe key elements behind the implementation
of \name\ and the reported proof-of-concept studies.
The complete implementation and all the proof-of-concept studies
are made available as open source software\cite{Spi23a},
and can be further examined or run on the openly available data
that \name\ uses to obtain additional details.

\subsection*{Implementation.} 
\name\ is designed around modules that handle the Crossref, ORCID, ROR, and
flat file (journal names, funders, open access journals) data sources.
Each module defines the data source's schema in terms of tables and
their columns.
The schema's SQL DDL implementation is used to define the corresponding
tables in a populated database.
The schema is also used for analyzing and satisfying vertical slicing
requests when populating databases with Crossref and ORCID data,
and for running Crossref queries without populating a database.
Where possible, data are on-the-fly decompressed, extracted from an
archive (with a tar or zip structure), and parsed in chunks,
thus avoiding the storage cost of the entire decompressed data set
(1 TB for Crossref and 467 GB for ORCID).

In its simplest form \name\ can evaluate an SQL query directly on
the Crossref dataset, often to perform exploratory data analysis.
Results can be saved as a CSV (comma-separated values)
file or iterated over through Python code
for further processing.
Both of these modes have limitations in terms of
performance,
aggregation of query results, and
combination of data from multiple sources.

In most cases \name\ is used to populate an
SQLite database\cite{Owe06}.
Multiple alternatives were considered for the data back-end.
The native Crossref tree format would suggest a JSON-style NoSQL
database.
Redis\cite{Car13} would be an efficient back-end, but would limit the
amount of data to that that could be processed in memory.
MondGB\cite{BBGV16} would address the capacity limitation, at the cost
of a more complex installation process due to licensing issues.
A client-server relational database system could also be used
to offer improved integrity constraints and more advanced
query optimization facilities.
However, these options require the installation, configuration,
connection, and maintenance of a separate database process,
which is not a trivial task,
especially for researchers outside the computing field.
Consequently, the adopted approach uses the SQLite embedded database,
which is
embedded into \name\,
directly available in Python,
easily installable as a command-line tool in all popular computing platforms,
and accessible through
APIs for diverse programming languages and environments.
Transferring a database between computers only involves
copying the corresponding file.
Consequently, there is no need to setup, configure, and maintain
a complex client-server relational or NoSQL
database management system.
Despite its -Lite suffix,
SQLite supports large parts of standard SQL\cite{ISO9075:2003}
(including window functions and recursive queries),
and employs sophisticated query optimization methods;
these feature made it ideal for use in \name.
SQLite's main downsides --- lack of multi-user and client-server support ---
are not relevant to common \name\ use cases.

Direct queries on the Crossref data set (without populating a database)
are implemented by defining SQLite virtual tables that correspond
to the offered schema through the Python {\em apsw} module.
Crossref data are distributed in the form of about 26 thousand
compressed containers.
Running a query on them is trivial when the query accesses a single table:
the virtual table implementation moves from one container to the next
as the table is scanned sequentially.

Direct Crossref queries involving multiple tables are handled differently.
First, a dummy query execution is traced to determine the tables and fields
it requires.
Then, in-memory temporary tables are populated with the required data
from each container,
and the query is executed repeatedly on the instantiated tables.
This approach works for all cases where relational joins
happen within a Crossref container
(e.g. works with their authors or references).
More complex cases, such as relational joins between works or aggregations,
require the population of an external database.

Database population with vertical slicing is implemented by attaching
the virtual database to the one to be populated and running
{\tt SELECT INTO} {\em populated-table} {\tt  SELECT FROM}
{\em virtual-table} queries.
A condition specifying a container-identifier is added to each query,
so that all tables for each decompressed container can be populated
before moving to the next container.
As before, query tracing is used to identify the tables and fields
that the user has asked to populate.

When the database population specifies horizontal slicing
(through a row-selection SQL expression)
the expression is evaluated sequentially on each container
through the following steps.
\begin{enumerate}
\item Create a query based on the specified expression and trace it
to determine the tables and fields required for evaluating the expression.
\item Create the empty tables to be populated.
\item Calculate the topological ordering\cite{Dia69} of the specified tables,
which will be used to join them in order to evaluate the expression.
\item Iterate through the Crossref containers doing the following.
\begin{enumerate}
\item For each table to be populated or participating in the SQL expression
create an in-memory mirror temporary table with the table's primary keys,
foreign keys, and fields participating in the query.
\item Create another temporary table made by joining the mirror tables,
evaluating the SQL expression as a {\tt WHERE} condition, and
inserting the resulting rows.
This table contains the identifiers of matched works.
\item Insert into each populated table records associated with the
matched works (directly or through {\tt JOIN} relations) selected from
the Crossref container being processed.
\end{enumerate}
\end{enumerate}

The database population design uses two techniques to improve its
performance.
First, records are bulk-inserted in batches by attaching directly
the virtual tables to the database to be populated, and by having
the database engine perform the insert operations with a single
internal command for each batch.
This avoids a round-trip cost of obtaining the data in \name\ and then
storing them back in the database.
Second, database indices over the containers in which the data are split
are implemented and used so as to access each file in turn for populating
all required tables.
The correspondingly improved locality of reference\cite{Den05} is then utilized
by caching the decompressed and parsed file contents.

The performant matching of author affiliations with RORs is based on multiple
applications of the Aho-Corasick string-matching algorithm\cite{FGREP}.
First, an Aho-Corasick automaton is created with all unique
organization names, aliases, and acronyms.
Second, the automaton is used to find entries in it that also match
other entries
(e.g. ``ai'', the acronym of the ``AI Corporation'', also matches
the organization name ``Ministry of Foreign Affairs''),
and mark them for removal to avoid ambiguous matches.
Finally, a new automaton constructed from the cleaned-up entries is
used to find the longest match associated with each author affiliation
string.
This is stored in the database as the affiliation's organization identifier.
When the \name\ user specifies that affiliations should match
the ultimate parent organization,
a recursive SQL query adds a ``generation'' number
to each matched organization,
an SQL window (analytic) function\cite{Zem12} orders results by generation,
and a final selection query obtains the ROR identifier
associated with the most senior generation.

\subsection*{Proof-of-concept studies framework.} 
We structured most proof-of-concept studies we presented
as a series of queries that build on each other.
This aids comprehensibility, testability, analysability, and recoverability.
We specified the corresponding workflow using Makefiles\cite{MAKE}
based on the {\em simple-rolap} system,
which manages relational online analytical processing tasks\cite{GS17,GS17b}.
The {\em simple-rolap} system establishes the dependencies between queries
and executes them in the required order.
Most studies start with a population phase, which fills a database
with the required horizontal and vertical data slices.
In many cases, we used the {\em rdbunit} SQL unit-testing framework
to test SQL queries\cite{GS17b}.
We employed a shared Makefile with rules and configurations
that satisfy dependencies required by more than one study.
For example, the common Makefile contains rules to download required data sets
and to populate the database with the datasets that do not support slicing
(journals, RORs, ASJCSs, DOAJ, and funders).
The {\tt common} directory where the shared Makefile resides
also hosts downloaded data sets to minimize useless data duplication.
All proof-of-concept studies can be executed by navigating to the
corresponding directory (under the {\tt examples} directory of the
\name\ source code distribution), and running {\tt make}.

In total the proof-of-concept studies distributed as examples with \name\
comprise
96 SQL query files amounting to 1\,788 (commented) lines of code.
The query operations are organized by 15 Makefiles (403 lines),
which populate the databases and execute the queries in the required order.
The queries create 37 intermediate tables and facilitate their
efficient operation through the creation of 42 table indices.
The charts, tables, and numbers in this report are based on
data produced by 60 SQL queries that obtain their results
from intermediate tables,
from populated databases, or
directly from the Crossref data set.

\subsection*{Research synthesis studies.} 
We calculated the numbers associated with research synthesis studies
by processing the output of \name\ run on the Crossref data with an SQL
query that matches specific words in publication titles.
The query's terms are structured to give precedence to the
characterization of titles indicating a systematic review as such,
classifying the rest as (unspecified) secondary studies.
In the Figure~\ref{fig:synthesis}
we combined the plotting of bibliometric (BM) and scientometric (SM)
studies,
and did not plot figures for mapping reviews (MR), umbrella reviews (UR),
and tertiary studies (S3) --- 6061 publications in total.
We also did not plot
the 400 identified studies published before 1971
as well as studies published after 2021,
as only part of the year 2022 data were available.
Note that works containing ``bibliometric'' or ``scientometric'' in their
title may either employ these methods\cite{CSMD21}
or refer to them\cite{Bra88}.
We used another query with the same terms to list the 30 studies published
before 1950 and obtain the earliest one\cite{BBF46} among them.

\subsection*{Crossref graph database.} 
We created a vertical slice of the complete Crossref database,
which we used for a number of purposes.
The slice contains mainly the primary and foreign keys
(including DOIs, and ORCIDs) of all entities,
plus the publication year, author affiliation names, and work
subjects, which are not normalized.
We also ran \name\ to populate the database with the
Scopus All Science Journal Classification Codes (ASJCs)\cite{SMWS17} and RORs,
and linked work subjects to ASJCs and author affiliations to RORs.

We used the graph database to extract most of the database contents
metrics provided in the main text and
in Tables~\ref{tab:crossref}~and~\ref{tab:orcid}.
This was done through simple {\tt SELECT Count(*) FROM table} or
{\tt SELECT Count(DISTINCT id) FROM table}
SQL queries.

\subsection*{Scientific publishing evolution.} 
In common with other studies\cite{WJU07,PLF23} we limited our examination
to works published after World War II, in order to avoid misleading
comparisons with the markedly different scientific and publication
environment that preceded it.
We calculated most numbers used for plotting Figure~\ref{fig:yearly-evolution}
from the populated Crossref graph database.
We obtained the number of published works and journals through
SQL {\tt Count} aggregations of the underlying data grouped by year.
We obtained the ratios of
authors per work and
references per work
by counting the corresponding elements of the associated detail
tables and then obtaining SQL {\tt Avg} aggregations grouped by year.
We joined papers with their citations using the document's DOI as a key
and used this to calculate
the two-year impact factor,
the received citations per work,
the twenty-year impact factor, and
the proportion of all published works cited at least once each year.
For calculating the last two
we used SQL window (analytic) functions\cite{Zem12} to obtain an accumulating sum and
a twenty-year rolling sum over the number of works published each year.

We calculated the number of pages per work by populating a database
with the required work and author details, and  by
extracting the starting
and ending page number from Crossref works that contain a dash in
their {\tt pages} field.
This process excludes single-page works reported with only a single
page number (rather than as a range with the same starting and ending page).
We excluded from the data records with a zero or negative number of pages or
those having more than 1000 pages, because the latter (164387 records)
were often derived
from data-entry mistakes, such as repeated page digits (e.g. {\tt 234-2366}),
as well as
unusable data formats, such as {\tt 1744-8069-5-32}.

We calculated a measure that can be used to track author productivity
(works per author) despite clashes in author names, by taking advantage
of the fact that we display productivity in relative terms (adjusting it
to be 1 in 1950).
In absolute terms authors with same names increase the
productivity's absolute value.
(An author named Smith, Kim, or Zhu would appear extremely prolific.)
However, assuming that the ratio of clashing names in the population does not
change, the effect of duplicate names on the relative productivity
measurements is cancelled out.

As an example consider that 50\% of all authors are named Smith,
all other authors have distinct names,
and in 1950 1000 authors write 2000 works.
The actual productivity should be 2.
By not distinguishing authors with clashing names
the situation will appear as 500 authors writing 2000 works, i.e. a
productivity of 4.
Consider now in 2020 10\,000 authors writing 80\,000 works.
The actual productivity should be 8.
The situation will however appear as 5\,000 authors
writing 80\,000 works, i.e. a
productivity of 16.
While the derived absolute productivity numbers are incorrect,
the ratio of the correct productivity numbers
is the same as the ones that do not take name clashes into account:
$2/8 = 4/16$.

Note that an actual study of author productivity would need to test and
control for the assumption we made, because the population's composition
might change over the years to include authors from ethnic backgrounds
with more or fewer common name clashes,
(for example, about 80\% of China’s population
shares the 100 most common Chinese surnames)\cite{Wan19}
making the phenomenon more frequent or less frequent over time.
Using \name\ we obtained frequently-occurring names at the two ends
of the examined period and found that in 1950 the five most frequent
names were
W. Beinhoff (161 works --- 0.10\% of the total names),
E. Rosenberg (149 --- 0.09\%),
F. De Quervain (115 --- 0.07\%),
G. Niemann (114 --- 0.07\%), and
A. Eichler (105 --- 0.06\%);
whereas in 2021 they were
Y. Wang (63363 --- 0.23\%),
Y. Zhang (57414 --- 0.21\%),
Y. Li (51792 --- 0.19\%),
Y. Liu (46013 --- 0.17\%), and
X. Wang (44805 --- 0.16\%).
The different percentages at the two periods' ends indicate that further
controls for this change would be required,
e.g. by measuring name clashes through ORCIDs.

We calculated the \cdv\ index\cite{FO17} of Crossref publications by
populating a database with their publication date and DOI,
as well as the DOIs of the corresponding references.
Given the available data, we were able to calculate the \cdv\ index
for six additional years (until 2016) compared to previous results\cite{PLF23},
using the remaining five complete years we had at out disposal (2017--2021)
to obtain the required citations' window.

The \cdv\ calculation proved to be computationally challenging.
The processing for the already-published 1945--2010 range,
taking into account even publications lacking reference data,
required more than five days of computing and about 40 GB of RAM.
Surprisingly,
extending the range to 2016 increased the required time to 45 days.
To address this we enhanced the original CD calculation algorithm implementation
to use more efficient data structures and algorithms converting it to C++\cite{FS23}.
We employed a union of pointers and integers to efficiently
represent vertices internally and in Python code, and used C++
sorted vectors and sets to improve memory allocation and searching
for nodes.
Furthermore, we rewrote the CD calculation process in C++
in order to parallelize it while maintaining in memory a single
copy of a 49 GB graph data structure.
(It turned out that this could not be done neither with Python's threads
nor with forked processes.)
The improvements in computational efficiency
allowed us to perform the \cdv\ index calculation in
33 hours of elapsed time using 40 hours of CPU time and 49 GB of main memory.

To allow other researchers to build on this data without incurring the
associated high computational cost,
we have made the resulting data set containing the DOI and the \cdv\ index
for 50\,937\,400 publications in the range 1945--2016
openly available\cite{Spi23b}.
This improves upon previously available data\cite{FPL22},
which extends to 2010 and only provides the time-stamp of each publication,
without other uniquely distinguishing publication identifiers.

\subsection*{Field dependencies.} 
We calculated strong dependencies between fields
(Figure~\ref{fig:subject-hierarchy})
by using work references and subjects in the Crossref graph database
to construct a table containing
the number of citations between fields.
Based on it we calculated for each field pair
its ``strength'' (sum of incoming and outgoing citations)
and its ``fundamentalness''
(ratio of a field's outgoing citations over the pair's strength).
In the plotted results we included the top 50 field associations
in terms of strength from the top 10\% in terms of fundamentalness.
We furthermore
excluded pairs associated with the ``Multidisciplinary'' subject
and also links within fields.
Both sides in the Figure represent
a small fraction of the total field citations,
and are drawn on different scales:
the outgoing citations shown amount to 0.8\% of the fields' total
and the incoming ones amount to 0.2\% of it.

\subsection*{Field evolution.} 
We calculated the evolution in the number of publications
across scientific fields (Figure~\ref{fig:fields-by-year})
by propagating the specific fields associated
with each work in the Crossref graph database to the more general
containing field.
For that we used as general fields the Scopus ASJCs that ended in ``00'',
and as their sub-fields those that started with the same numeric prefix.
For example, we allocated publications under the subject of ``Catalysis'' (1503)
to ``General Chemical Engineering'' (1500).
We then calculated total publications in 1945 and 2021,
changes in the percentages of a field's publications in terms of the
total at the two time points,
and included in the Figure the ten fields with the largest change
whose publications amounted to more than 2\% of the 2021 total.

\subsection*{COVID-19 metrics.} 
To study COVID-19 publications we populated a database
with a full horizontal slice of the Crossref data by
specifying as the row selection criterion
works containing the string ``COVID'' in their title or abstract
(``covid'' is not part of any English word).
We also linked works to their subjects and author affiliations
to the corresponding RORs.
We obtained organizations publishing COVID-19 research by assigning
author affiliations to works, and by counting both ROR-matched
affiliations and unmatched affiliations as simple text.

\subsection*{Number of COVID-19 study authors.} 
We calculated the approximate number of researchers who worked on
all COVID-19 studies
by starting with the number of unique (author given-name, author surname) pairs
in the set of all COVID study authors $N_{an}$.
The number of true authors could be higher if many authors share the same name
or lower if the same author appears differently
(e.g. through the use of initials) in some publications.
We address this by obtaining from the set of authors with an ORCID
the number of distinct ORCIDs $N_o$, which is the true number of authors
in that set,
and the number of distinct names $N_{on}$, which approximates any bias
also found in $N_{an}$.
We then consider the true number of authors as
\[
N_{an} \frac{N_o}{N_{on}}
\]

\subsection*{Journal impact factor.} 
We calculated the 2021 journal impact factor\cite{Gar06} by populating a database
with the keys, ISSNs, publications years, and pages of works and
their references published between 2019 and 2021.
We then created a table associating works with journal ISSNs.
From this table we obtained citations published in
2021 to works published in 2019--2020
(the impact factor's numerator).
We further filtered works to identify ``citable'' items,
which Clarivate defines as those that make a substantial contribution to
science and therefore do not include elements such as editorials and letters.
For this we used as a rough heuristic works longer than two pages.
(We also included works lacking a page range.)
From the count of citable items per journal we obtained the number of publications published in the 2019--2020 period (the denominator).
Finally, to compare our results with the numbers published by Clarivate
we associated each impact factor metric with all ISSNs known for
a journal (electronic, print, alternative), excluding the ``alternative''
ISSNs of one journal used as primary for another journal.

\subsection*{Productivity metrics.} 
We obtained the h5-index\cite{Hir05} productivity metrics by populating databases
with data sliced vertically to include the keys of works and references
and horizontally to include items published in the period 2017--2021.
For the software engineering venue metrics we selected the
examined conferences based on the DOI prefix assigned to the conference
publications each year.
(In retrospect, we could have used the container title.)
For each entity we counted its citations and then used an SQL
window (analytic) function to partition the results over the entity's
key (ISSN, ORCID, or conference acronym),
number each set's rows,
and select only those with a rank lower or equal to the
corresponding number of citations.

To study the citation graph of top-ranked authors we obtained
a) a random sample of 50 works written by top-ranked
authors, and
b) a random sample of 50 works from all other publications,
paired with the ones selected from the top-ranked authors to have the same
number of citations as them.
For each publication in the two samples, we created a separate graph
containing the work $w$,
the set $S$ of the works $w$ cites and the works that cite $w$,
and then again the set $S'$ of the works $w'$ that cite or are
cited by $w \in S$.
The graph's edges are citations from one work to another.
We created each citation-induced graph with a Python program
querying the populated database and employed the {\em NetworkX}\cite{HSS08}
Python package to calculate the graph's average clustering.

SQLite lacks the ability to provide a seed to its {\em Random()} function,
which is required for obtaining random samples in a deterministic manner.
We worked around this limitation by multiplying the identifier of each author
(which is sequentially allocated and therefore not random) with a seed value,
and used the last decimal digits of the result
to place each work in a pseudo-random ordering,
which we used to obtain the required works.

\subsection*{Performance and map of available code.} 
{\small
\begin{table}
\caption{\label{tab:perf}Performance Summary}
\begin{center}
\begin{tabular}{lcrrrrrrrr}
\toprule
Task	& H	& $S_{pop}$	& $T_{pop}$	& $S_{ROLAP}$	& \# Q		& $T_{Qtot}$	& $T_{Qmean}$	& $T_{Qmax}$ \\
\midrule
author-productivity	& R	& 18 GiB	& 4:23:09	& 12 KiB	& 4	& 1:50:49	& 0:27:42	& 0:55:55 \\
cdindex	& S	& 90 GiB	& 10:59:08	& 3 GiB	& 3	& 2:49:09	& 0:56:23	& 2:23:44 \\
covid	& R	& 3 GiB	& 8:05:45	& —	& 6	& 0:29:29	& 0:04:54	& 0:15:15 \\
crossref-standalone	& T	& —	& —	& —	& 3	& 11:50:21	& 3:56:47	& 3:58:36 \\
graph	& R	& 189 GiB	& 7:30:53	& 283 MiB	& 37	& 27:55:29	& 0:45:17	& 11:57:56 \\
impact-factor-2021	& T	& 30 GiB	& 7:18:05	& 2 GiB	& 12	& 2:56:54	& 0:14:44	& 0:45:45 \\
journal-h5	& S	& 33 GiB	& 20:49:03	& 5 GiB	& 4	& 2:06:04	& 0:31:31	& 0:58:58 \\
orcid	& T	& 6 GiB	& 3:49:02	& —	& 1	& 0:02:02	& 0:02:02	& 0:02:02 \\
person-h5	& T	& 50 GiB	& 8:31:05	& 5 GiB	& 7	& 3:15:12	& 0:27:53	& 0:47:47 \\
research-synthesis	& S	& —	& —	& —	& 1	& 7:01:52	& 7:01:52	& 7:01:52 \\
soft-eng-h5	& T	& 137 MiB	& 9:51:14	& 7 MiB	& 5	& 0:01:01	& 0:00:12	& 0:01:01 \\
yearly-numpages	& T	& 4 GiB	& 4:08:07	& 2 GiB	& 6	& 2:54:52	& 0:29:08	& 0:56:56 \\

\bottomrule
\addlinespace
\multicolumn{9}{@{}l@{}}{H(ost) R: Intel i7-10700 CPU @ 2.90GHz, 16 MiB cache, 32 GiB DDR4 RAM, SSD storage;}\\
\multicolumn{9}{@{}l@{}}{H(ost) S: Intel E5-1410 CPU @ 2.80GHz, 10 MiB cache, 72 GiB DDR3 RAM, magnetic disk;}\\
\multicolumn{9}{@{}l@{}}{H(ost) T: Intel i7-7700 CPU @ 3.60GHz, 8 MiB cache, 16 GiB DDR4 RAM, magnetic disk;}\\
\multicolumn{9}{@{}l@{}}{$S_{pop}$: Populated database size; $T_{pop}$ Time to populate database;}\\
\multicolumn{9}{@{}l@{}}{$S_{ROLAP}$: Relational analytical processing database size; \# Q: number of queries run;}\\
\multicolumn{9}{@{}l@{}}{$T_{Qtot}$: total query time; $T_{Qmean}$: average (arithmetic) query time; $T_{Qmax}$: maximum query time}\\
\end{tabular}
\end{center}
\end{table}
}

Performance details of the tasks reported here are
summarized in Table~\ref{tab:perf}.
The reported figures are associated with two parts of each analysis.
First, the population of an SQLite database with a horizontal and
vertical slice of Crossref (and other) data.
The size of this database (together with any indexes created in it)
is reported as $S_{pop}$ and the corresponding time required to
populate the database as $T_{pop}$.
Second, the execution of ROLAP SQL queries to obtain the required results.
In many cases the queries generate secondary tables,
which are stored in a separate ROLAP (analysis) database.
The size of the ROLAP database is reported as $S_{ROLAP}$
and the time required to run the queries as  $T_{Qtot}$,
$T_{Qmean}$, and $T_{Qmax}$.
The query run time includes time for creating indexes,
intermediate tables, and final reports.
For tasks where no population figures are listed,
the queries are run directly on the Crossref containers.
For tasks where no ROLAP size is listed,
the queries generate the results directly from the populated database, 
without creating any intermediate tables.
All reported times are elapsed (wall clock) times
shown in hours:minutes:seconds format.

The numbers shown on the table involve executions on idle or lightly
loaded hosts with processes taking up a single core.
The memory use (maximum resident set size --- RSS) of population and query tasks
averaged 533 MiB
(minimum 4 MiB, maximum 14 GiB, median 21 MiB, $\sigma = 1388921$).

In addition to the SQL query times reported in the table,
the author graph analysis was performed by a Python program
(0:08:38 elapsed time, 55 MiB RSS)
and the \cdv\ calculation by a multithreading C++ program
(9:24:20 elapsed time on 8 cores, 66:48:37 CPU time, 49 GiB RSS).

The task names in Table~\ref{tab:perf}
mirror the contents of the \name\ source code distribution
{\tt examples} directory.
They are associated with what is reported in this work as follows.
The {\em author-productivity} and {\em yearly-numpages} tasks were used
to derive the corresponding yearly evolution lines shown in
Figure~\ref{fig:yearly-evolution}.
The {\em crossref-standalone} task derived the evolution in the number
of journals (Figure~\ref{fig:yearly-evolution}),
the yearly availability of abstracts
(Figure~\ref{fig:yearly-availability}),
and the types of works available in Crossref.
The {\em graph} task was used to derive
the remaining yearly evolution and yearly availability metrics,
the Crossref record metrics (Table~\ref{tab:crossref}),
as well as the relationships between scientific fields
and the evolution of publications in them
(figures~\ref{fig:subject-hierarchy} and~\ref{fig:fields-by-year}).
The {\em orcid} task derived the ORCID metrics shown in
Table~\ref{tab:orcid}.
The {\em research-synthesis} task derived the evolution in
systematic literature reviews (Figure \ref{fig:synthesis}).
Finally,
{\em cdindex} derived the \cdv\ index,
{\em covid} the COVID-19 figures,
{\em impact-factor-2021} the 2021 journal impact factor,
{\em journal-h5} the journal h5-index,
{\em person-h5} the person h5-index, and
{\em soft-eng-h5} the h5-index of software engineering venues.

\subsection*{Statistical analysis.} 
For reporting the correlation between the metrics obtained by
\name\ and existing ones and for comparing the graph clustering
coefficients between two populations we used the functions
{\em spearmanr} and {\em mannwhitneyu} from the Python package
{\em scipy.stats}.
All calculations were performed with ``two-sided'' as the
alternative hypothesis (the default).
No other options were provided to the function calls.
For the analysis and charting we used
Python 3.9.10 with
matplotlib 3.3.4,
numpy 1.20.1,
pandas 1.2.3,
pySankey 0.0.1, and
scipy 1.6.2.

\newpage

\bibliography{macro,classics,ddspubs,mybooks,myart,various,unix,alexandria3k,coderead,isostd} 

\section*{Acknowledgments} 
The author thanks
 Panos Louridas,
 Arie van Deursen,
 Theodoros Evgeniou,
 Alberto Bacchelli,
 Dirk Beyer,
 and
 Dimitris Karlis for valuable advice and feedback.

\begin{description}
\item[Competing interests:]
The authors declares that he has no competing interests.

\item[Data and materials availability:]
A versioned release of the source code of \name\ and the proof-of-concept
examples presented in this study is available on Zenodo\cite{Spi23a},
licensed under the GNU General Public License v3.0.
A replication package with the results data and Python scripts used
for the statistical analysis and charting
of the reported example studies is also available
on Zenodo\cite{REPLICATION}.
Current versions of \name\ are made available
for installation through PyPi \url{https://pypi.org/project/alexandria3k/}
and
for contributions, feature requests, and issue reporting
on GitHub \url{https://github.com/dspinellis/alexandria3k}.
\end{description}
The data used in the example studies are available as follows.

{\small
\begin{tabular}{ll}
Crossref Apr. 2022 Public Data	File	& \surl{doi:10.13003/83b2gq} \\
ORCID Public Data File 2022 v. 4	& \surl{doi:10.23640/07243.21220892.v4} \\
ROR Data v1.17.1			& \surl{doi:10.5281/zenodo.7448410} \\
Open access journals	& \surl{https://doaj.org/csv} \\
Funders			& \surl{https://doi.crossref.org/funderNames?mode=list} \\
Journals		& \surl{http://ftp.crossref.org/titlelist/titleFile.csv} \\
\end{tabular}
}

The 2021 Journal Impact Factor data used for assessing the numbers
obtained by \name\ are available from Clarivate,
but restrictions apply to the availability of these data,
which were used under license for the current study,
and so are not publicly available.
These data are however available from the authors upon reasonable
request and with permission of Clarivate.

\newpage

\ifx\review\relax
\section*{Supplementary figures}
\renewcommand{\thefigure}{S\arabic{figure}}
\renewcommand{\figurename}{Supplementary Figure}
\else
\section*{Appendix: Database schemas}
\fi

\begin{figure}
\includegraphics[width=\columnwidth]{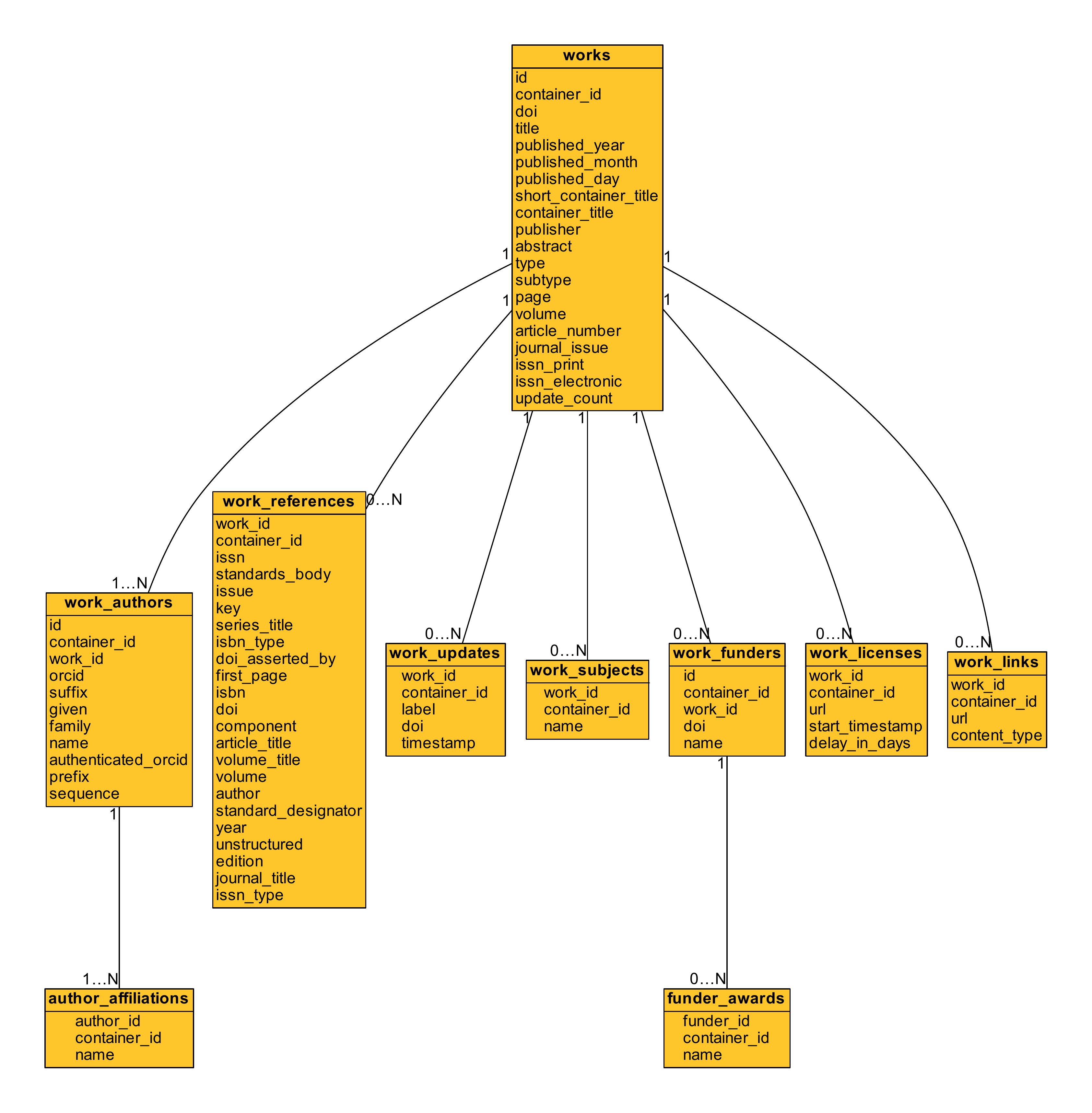}
\caption{\label{fig:schema-crossref}Relational schema of Crossref tables
in a fully-populated database.}
\end{figure}

\begin{sidewaysfigure}
\includegraphics[width=\columnwidth]{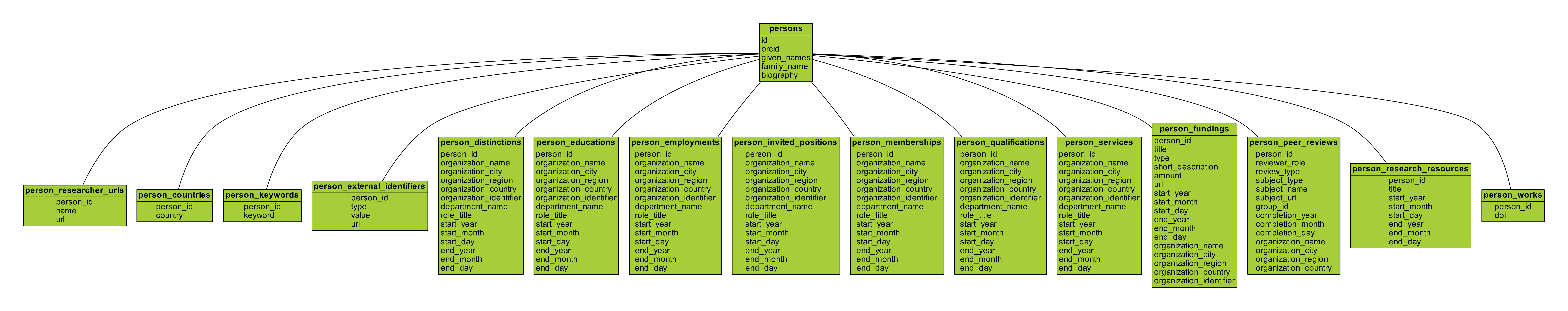}
\caption{\label{fig:schema-orcid}Relational schema of ORCID tables
in a fully-populated database.}
\end{sidewaysfigure}

\begin{figure}
\includegraphics[width=\columnwidth]{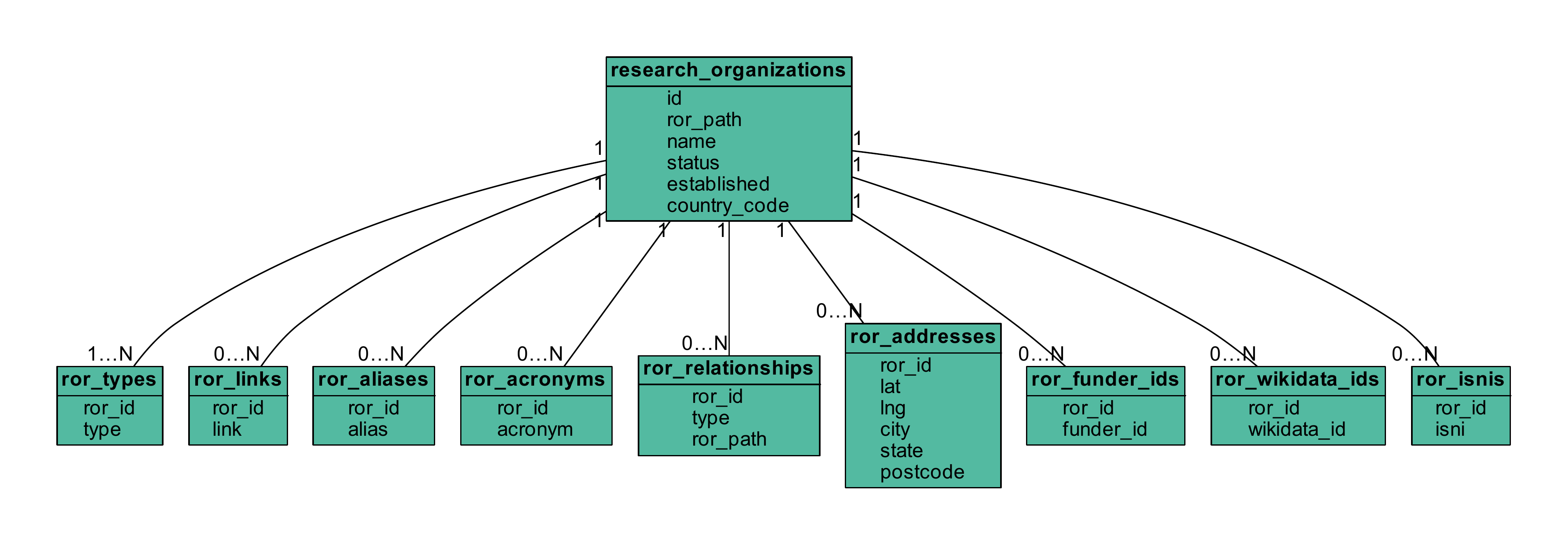}
\caption{\label{fig:schema-ror}Relational schema of ROR tables
in a fully-populated database.}
\end{figure}

\begin{figure}
\includegraphics[width=\columnwidth]{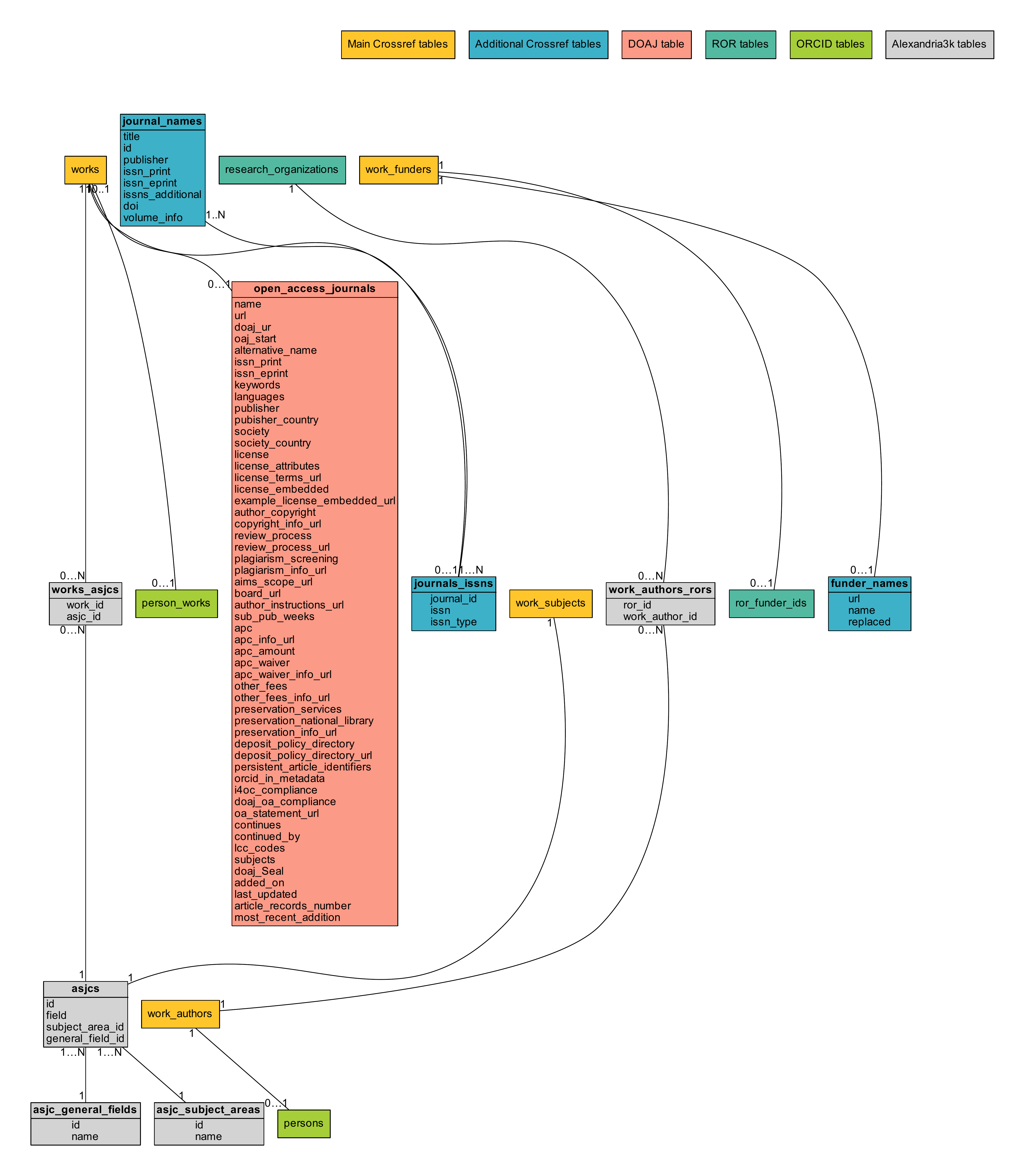}
\caption{\label{fig:schema-other}Relational schema of other tables
and \name-generated links
in a fully-populated database.}
\end{figure}

\end{document}